\documentclass[amstex,aps,preprint,superscriptaddress]{revtex4-2}%
\usepackage{amsmath}
\usepackage{amsfonts}
\usepackage{amssymb}
\usepackage{graphicx}
\usepackage{xcolor}%

\begin{document}

\title{Dynamics, locality and weak measurements: trajectories and which-way
information in the case of a simplified double-slit setup}
\author{F.\ Daem}
\author{A. Matzkin}
\affiliation{Laboratoire de Physique Th\'eorique et
	Mod\'elisation, CNRS Unit\'e  8089, CY Cergy Paris
	Universit\'e , 95302 Cergy-Pontoise cedex, France}

\begin{abstract}
Understanding how the interference pattern produced by a quantum particle in
Young's double-slit setup builds up -- the \textquotedblleft only
mystery\textquotedblright\ of quantum mechanics according to Feynman -- is
still a matter of discussion and speculation. Recent works have revisited the
possibility of acquiring which-way information based on weak
measurements.\ Weak measurements preserve the interference pattern due to
their minimally perturbing character while still leading to a final position
detection. Here we investigate a simplified double-slit setup by including
weakly coupled pointers. We examine how the information provided by the weak
pointers can be interpreted to infer the dynamics within a local picture
through \textquotedblleft weak trajectories\textquotedblright. We contrast our
approach with non-local dynamical accounts, such as the modular momentum
approach to weak values and the trajectories defined by the de Broglie-Bohm picture.

\end{abstract}
\maketitle
\newpage

\section{Introduction}

In the \emph{Character of Physical Law} \cite{CPL}, Feynman discusses at
length the double slit experiment. He asks whether it is true that
\textquotedblleft electrons must go either through one hole or
another\textquotedblright. His answer is that if we look we will indeed find
the electron at one slit or the other, but we will then loose the interference
pattern.\ If we do not look, such a claim would be an \textquotedblleft
error\textquotedblright. Since then the weak measurement (WM) scheme, a
minimally-disturbing process that enables one to gain information on the
property of a quantum system at an intermediate time $t$ as the system evolves
from an initially prepared state $|\psi(t_{i})\rangle$ to final state
$|\chi(t_{f})\rangle$ has been proposed \cite{AAV}. Although they are
difficult to undertake experimentally, weak measurements have been observed,
including in situations \cite{urbasi-exp} calling for Feynman's which slit
(or, more generally, which-path) question. Such techniques open an
observational window on the evolution of a quantum system between preparation
and final detection.

Several recent works \cite{mori1,narducci,luis,mori2,aharonov2S,hiley,PRA2022,hofmann}
have employed weak measurements in the context of the double slit setup. One
of the aims is to answer the which slit question, or rather more
precisely to account for the slit traversal dynamics. One series of works
\cite{mori1,mori2,narducci,PRA2022} focused on \textquotedblleft weak
trajectories\textquotedblright\ \cite{tanaka,matzkinWT,hiley}, that is
trajectories that are defined by a series of weak measurements of the system's
position. These weak trajectories are related \cite{matzkinPR} to the sum over the Feynman
paths of the system, so that one observes a superposition of
trajectories going through each slit, with the constraint that these
trajectories must be compatible with the post-selected quantum state at the
detection point on the screen. Other works \cite{aharonov2S,hofmann} have focused
more directly on the which-way question, that is whether by performing a weak
measurement one can tell that an electron went though a given slit while
maintaining interference at the detection screen. Employing a simplified setup
taking into account the sole dimension along the slits (or equivalently the
detection plane), Ref. \cite{aharonov2S} argues that it is possible to infer
which slit the particle has gone through without destroying the interference pattern,
while Ref. \cite{hofmann} argues, in line with the weak trajectories based results,
that the particle is delocalized on both slits.

In this paper, we study the simplified setup employed in Ref. \cite{aharonov2S}
with the tools introduced to investigate the full double slit setup in an
earlier work by one of us \cite{PRA2022}, work whose aim was precisely to
observe the superposition of paths coming from each slit at the detection
screen. In the full, two-dimensional setup, a great number of interfering
paths typically contribute to the weak values and to the final detection,
rendering the analysis convoluted. In the simplified setup, the
one-dimensional character of the problem makes it simple to capture the
dynamics in terms of weak pointers interacting locally with the system and
picking up the interference between two wavepackets. 

The paper is organized as follows.\ We will first briefly review wavepacket
propagation and the \textquotedblleft weak trajectories\textquotedblright%
\ resulting from the interaction between the wavepacket and a series of weakly
coupled pointers placed along the way (Sec. \ref{sec-2}). We will then
introduce in Sec. \ref{sec-simplified} the simplified double-slit setup and
derive the three types of weak values that may appear in this setup. Sec.
\ref{sec-inferring} will discuss how the dynamics can be inferred from these
weak values.\ We will contrast the resulting local dynamics from the one
supported by a non-local approach to the weak values \cite{aharonov2S,modular}
based on a localized particle subjected to non-local potentials. We will also
discuss the dynamical picture arising from the de Broglie-Bohm model \cite{holland,BH}, that is
well-known to be non-local; to this end, Bohmian trajectories in the presence
of pointers will be computed numerically. Our concluding remarks will be given
in Sec. \ref{sec-conc}.

\begin{figure}[ptb]
\centering \includegraphics[scale=1.0]{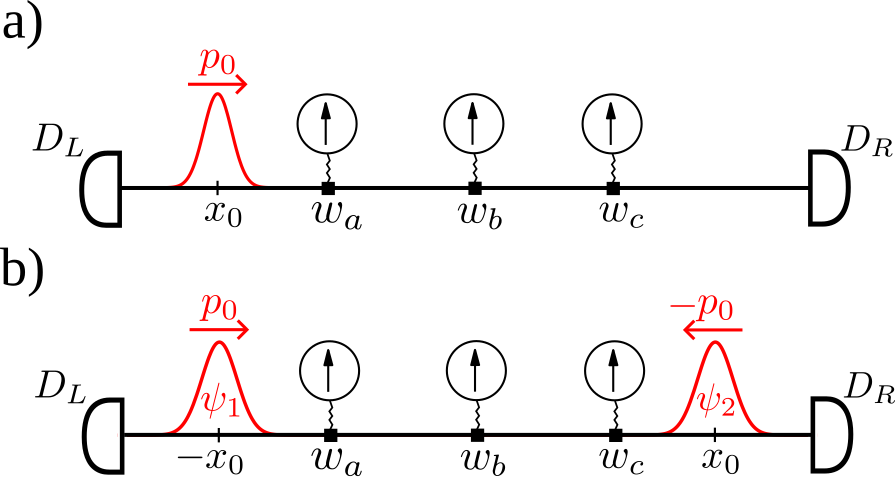}
\caption{(a) Schematics of a  
	wavepacket with initial mean position $x_0$ launched towards the detector 
	$D_R$. The wavepacket will be detected by $D_R$ with certainty
	interacting along the way with weakly coupled pointers such as those positioned at $w_a,w_b,w_c$. The shifts of these
	pointers define the ``weak trajectory'' of the particle. (b) Simplified double-slit setup: 
	The initial state is a superposition of two wavepackets $\psi_1$ and $\psi_2$ initially centered
	 at $-x_0$ and $x_0$ resp. (mimicking the positions of the slits), and with initial momentum $p_0$ and $-p_0$. Post-selection takes
	 place at $D_R$.}%
\label{fig-simple}%
\end{figure}

\section{Weak values and detection of a propagating wavepacket \label{sec-2}}

\subsection{Wavepacket evolution}

Let us consider the following simple situation: a wavepacket initially $(t=0)$
centered at $x_{0}$ with mean momentum $p_{0}$ propagates freely towards the
right $\left(  p_{0}>0\right)  $.\ Two detectors $D_{L}$ and $D_{R}$ are
placed respectively at $x_{L}$ and $x_{R}$ lying to the left and to the right
of the wavepacket's initial position (see Fig. \ref{fig-simple}(a)). For
definiteness we will choose the wavepacket $\psi_{G}(x,t)$ of the particle to
be a Gaussian of width $d$, given under free Hamiltonian evolution by
\begin{equation}
\psi_{G}(x,t;x_{0},p_{0})=\frac{\left(  \frac{2}{\pi}\right)  ^{1/4}\sqrt{dm}%
}{\left(  2d^{2}m+i\hbar t\right)  ^{1/2}}\exp\left(  -\frac{m(x-x(t))^{2}%
}{4d^{2}m+2i\hbar t}+\frac{ip_{0}\left(  x-x(t)\right)  }{\hbar}+\frac
{ip_{0}^{2}t}{2\hbar m}\right)  \label{gauT}%
\end{equation}
where $x(t)=x_{0}+p_{0}t/m$ is the position of the center of the wavepacket
($m$ is the particle mass). Eq. (\ref{gauT}) is obtained by applying the
particle's evolution operator $\hat{U}(t^{\prime},t)=e^{-i\frac{\hat{p}^{2}%
}{2m\hbar}(t^{\prime}-t)}$ to $\left\vert \psi_{G}((t=0)\right\rangle $.\ In the
position representation, the evolution operator is given by the so called free
propagator,%
\begin{equation}
K(x^{\prime},t^{\prime};x,t)\equiv\left\langle x^{\prime}\right\vert \hat
{U}(t^{\prime},t)|x\rangle=\left(  \frac{m}{2i\pi\hbar(t^{\prime}-t)}\right)
^{\frac{1}{2}}\hspace{0.05cm}e^{\frac{im(x^{\prime}-x)^{2}}{2\hbar(t^{\prime
}-t)}}\label{amplitudelibre}%
\end{equation}
Once the wavepacket is launched, and if the wavepacket broadening is deemed to
be negligible, the detector $D_{R}$ will click with certainty.

\subsection{From weak values to weak trajectories}

Let us now insert a pointer along the way between $x_{0}$ and $x_{R}$, say at
$x=w_{1}$ and let us assume the coupling between the pointer and the particle
is weak.\ More precisely, we will couple the probe momentum $\hat{P}$ to the
particle observable $\hat{A}$. This implies that in addition to the particle's
free Hamiltonian, there is an interaction term of the form
\begin{equation}
\hat{H}_{int}=g(t)\hat{A}\hat{P}f_{w}.
\label{hint}
\end{equation}
$g(t)$ determines the timing at which the coupling is turned on, a function
that will be non-vanishing only in a small interval centered on the
interaction time $t_{w}$. $f_{w}$ reflects the short-range character of the
interaction that is only non-vanishing in a small region near $w_{1}$. Since
we are assuming the interaction is weak, the evolution due to $\hat{H}_{int}$
in the propagator is taken only to first order \cite{matzkinPR}.\ The system
wavefunction $\psi_{G}$ is thus only slightly perturbed and continues to
evolve until at time $t_{f}$ a standard measurement is made, leaving the
system in a state $\left\vert \chi_{f}(t_{f})\right\rangle $ which is an
eigenstate of the measured observable. The initial state of the pointer
$\left\vert \phi_{i}\right\rangle $ becomes at $t_{f}$ $\left\vert \phi
_{f}\right\rangle =e^{-igA^{w}\hat{P}}\left\vert \phi_{i}\right\rangle $ where
$A^{w}$ is the weak value \cite{AAV} given here by%
\begin{equation}
A^{w}=\frac{\left\langle \chi_{f}(t_{f})\right\vert \hat{U}(t_{f},t_{w}%
)\hat{A}\hat{U}(t_{w},t_{i})\left\vert \psi_{G}(t_{i})\right\rangle
}{\left\langle \chi_{f}(t_{f})\right\vert \hat{U}(t_{f},t_{i})\left\vert
\psi_{G}(t_{i})\right\rangle }%
\end{equation}
where $g=\int dt^{\prime}g(t^{\prime})$  (see \cite{matzkinFP} for a
derivation employing the present notation). The final probe state is hence the
initial one shifted by $g\operatorname{Re}(A^{w})$.

The weak value is defined provided the post-selected state $\left\vert
\chi_{f}(t_{f})\right\rangle $ overlaps with the freely evolved particle state
at time $t_{f}$ in line with the idea that the interaction is so weak as not
to affect the detection probabilities.\ Here, let us set the pointer to couple to the
spatial projector at $w_{1},$ $\hat{A}=\left\vert x_{1}\right\rangle
\left\langle x_{1}\right\vert $, $D_{R}$ will click almost with certainty
provided  post-selection captures exactly the (unperturbed) system
wavefunction.\ \ This is tantamount to choosing $\left\vert \chi_{f}%
(t_{f})\right\rangle =\left\vert \psi_{G}(t_{f})\right\rangle $ and the weak
value is simply given by $\left\vert \psi_{G}(x_{w},t_{w})\right\vert ^{2}.$
More realistically, $\hat{A}$ should be taken as a projector over a finite
width \cite{oriols} and the weak value would be given by $A^{w}=\int_{f_{w}%
}\left\vert \psi_{G}(x^{\prime},t_{w})\right\vert ^{2}dx^{\prime}$ where the
integration is over the domain defined by $f_{w}$. Finally, consider placing a
series of identical weakly coupled pointers along the path from $x_{0}$ to
$D_{R}$, choosing the time window at which the coupling is turned on. By
tuning the time widow of each pointer so that it corresponds to the wavepacket
going through the position of the pointer, one can observe -- in principle
experimentally, by measuring the pointers' shift -- the \textquotedblleft weak
trajectory\textquotedblright\ \cite{matzkinWT} of the system as it moves from
$x_{0}$ to $D_{R}$.

Overall this gives a simple dynamical picture: recalling that the propagator
$K$ of Eq. (\ref{amplitudelibre}) can be written as a sum over the classical
paths connecting $\left(  x,t\right)  $ to $\left(  x^{\prime},t^{\prime
}\right)  $ -- the term $m(x^{\prime}-x)^{2}/2(t^{\prime}-t)\ $\ is the
classical action --, we see that the shift in each pointer can be understood
as the result of the interaction between each pointer and the particle
wavepacket propagating along the Feynman path connecting $x_{0}$ and $D_{R}$.

\begin{figure}[ptb]
	\centering \includegraphics[scale=1.]{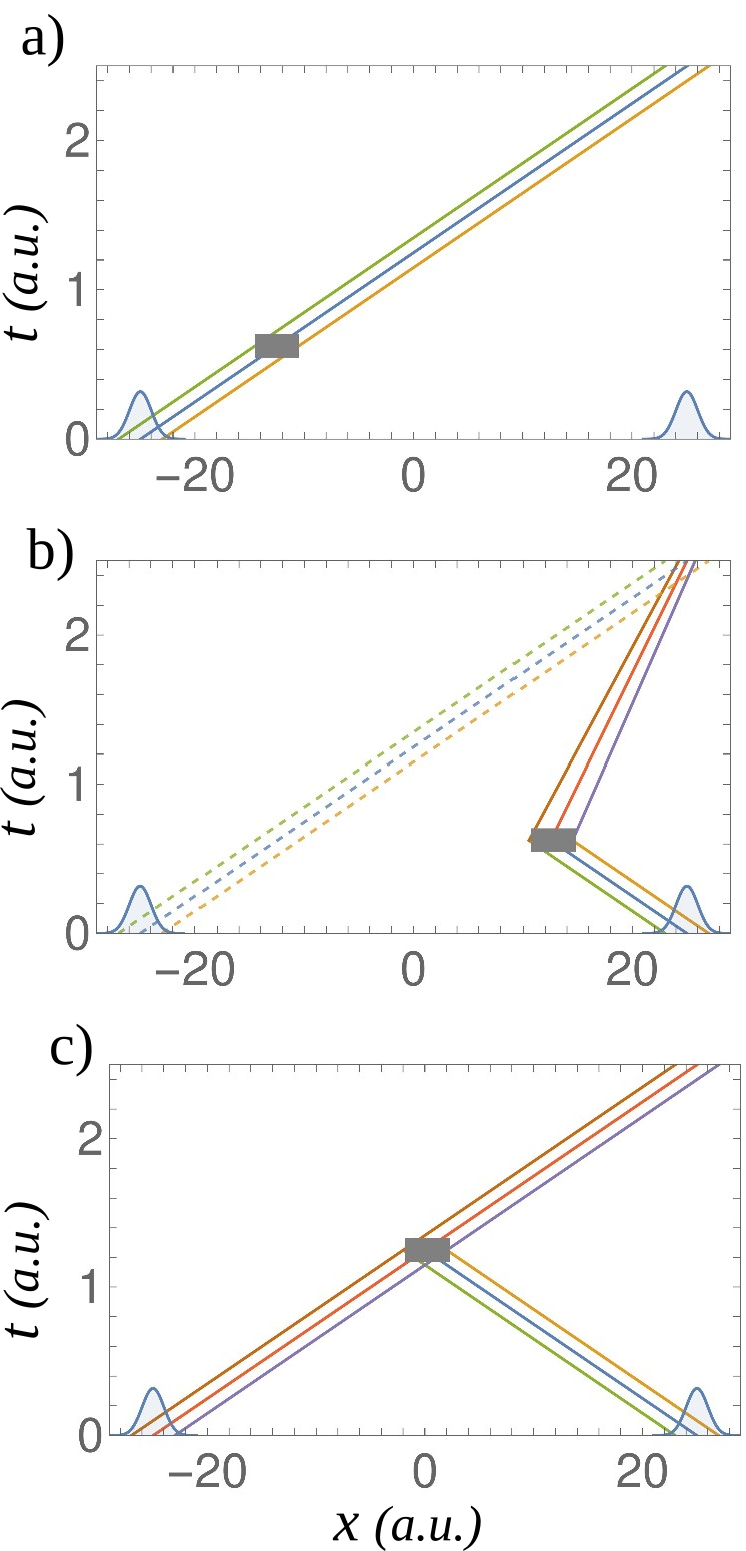}\caption{The three type of weak values mentioned in the text pictured
		with the corresponding weak trajectories; post-selection takes
		place at $D_R$ (we use atomic units and the system has unit mass).  (a) 
		The pointer placed at $w_a$ (gray box) interacts with the wave-packet initially centered at $-x_0$ (on the left) that will be postselected at $t_f$. (b) The weakly coupled pointer at $w_c$ interacts with the other wavepacket (originating from the right). The corresponding weak value will be non-zero if there are weak trajectories linking $w_c$ to the post-selection region at $D_R$. (c) The weak value at $w_b$ results from the superposition of both wavepackets. }%
	\label{fig-wt}%
\end{figure}

\section{Weak values and post-selection in a simplified
double-slit\label{sec-simplified}}

\subsection{The simplified double slit setup}

Let us now take the scheme of the preceding section but choose instead the
initial state as the superposition of two wavepackets initially on opposite
sides and launched in opposite directions,%
\begin{align}
\psi(x,t=0)= &  \frac{1}{\sqrt{2}}\left(  \psi_{G}(x,0;-x_{0},p_{0})-\psi
_{G}(x,0;x_{0},-p_{0})\right)  \label{supp}\\
\equiv &  \frac{1}{\sqrt{2}}\left(  \psi_{1}(x,0)-\psi_{2}(x,0)\right)
\end{align}
This can be seen as the part of the wavefunction in the double-slit problem
evolving in the plane of the slits.\ Recall indeed that the double-slit setup
takes place in a two dimensional plane, but the wavefunction in the direction
orthogonal to the detection plane is separable to a very good approximation
\cite{beau,PRA2022}. While Eq. (\ref{supp}) does not fully capture the
dynamics of the two-slit problem, it pinpoints in a simpler way the
\textquotedblleft which slit\textquotedblright\ question: if the wavepacket is
found say at $D_{R}$, can we say that this corresponds unambiguously to the
part of the wavepacket $\psi_{1}$ that was initially located on the right (at
$x_{0}$) ?

\subsection{Weak values and post-selection}

Consider Fig.\ \ref{fig-simple}(b), showing the position of 3 weakly coupled
pointers $w_{a}$, $w_{b}$ and $w_{c}$ and let $t_{f}$ denote the average time
it takes the wavepackets to reach a detector from its initial position. We
post-select on $D_{R}$ at $t=t_{f}$ in all the cases -- the post-selection
hence involves only $\psi_{1}$. We will see that the weak values involve
either a single wave-packet (which can only be $\psi_{1}$ given our
post-selection choice) or a combination of $\psi_{1}$ and $\psi_{2}$.

Assume we turn on the pointer $a$ in a time-window centered on $t_{1}\simeq
t_{f}/4$ when $\psi_{1}$, coming from the left, overlaps with $a$. The weak
value is
\begin{equation}
\Pi_{a}^{w}(t_{1})=\frac{\left\langle \chi_{R}(t_{f})\right\vert \hat{U}%
(t_{f},t_{1})\left\vert w_{a}\right\rangle \left\langle w_{a}\right\vert
\left.  \psi_{1}(t_{1})\right\rangle }{\left\langle \chi_{R}(t_{f})\right\vert
\left.  \psi_{1}(t_{f})\right\rangle }.\label{typ1}%
\end{equation}
For the moment the only assumption on $\left\vert \chi_{R}(t_{f})\right\rangle
$ is that it is localized at the right detector $D_{R}$. The weak value
(\ref{typ1}) involves solely the $\psi_{1}$ part of the wavepacket
(\ref{supp}), given that only $\psi_{1}$ interacts with $a$ at time $t_{1}$
($\psi_{2}(w_{a},t_{1})$ vanishes) and is detected at $D_{R}$.\ Therefore, on
this basis one might conclude that the particle detected by $D_{R}$ comes from
the wavepacket initially localized at $x=-x_{0}$ near $D_{R}$. The Feynman paths
contributing to $\Pi_{a}^{w}(t_{1})$ -- the weak trajectories -- are computed in
Fig. \ref{fig-wt}(a); these paths are the classical trajectories, reaching the weak pointer 
at $t=t_f/4$ and the post-selection region at $t_f$.

However, consider turning on the pointer $c$ at the same time $t_{1}$, when
the part $\psi_{2}$ of the wavepacket goes through it. The weak value is
\begin{equation}
\Pi_{c}^{w}(t_{1})=\frac{\left\langle \chi_{R}(t_{f})\right\vert \hat{U}%
(t_{f},t_{1})\left\vert w_{c}\right\rangle \left\langle w_{c}\right\vert
\left.  \psi_{2}(t_{1})\right\rangle }{\left\langle \chi_{R}(t_{f})\right\vert
\left.  \psi_{1}(t_{f})\right\rangle }.\label{typ2}%
\end{equation}
Now both $\psi_{1}$ and $\psi_{2}$ contribute to the weak value, $\psi_{2}$
through the interaction with $c$ and $\psi_{1}$ by the post-selection
condition.\ Note that for typical choices of $\left\vert \chi_{R}%
(t_{f})\right\rangle $, the term $\left\langle \chi_{R}(t_{f})\right\vert
\hat{U}(t_{f},t_{1})\left\vert w_{c}\right\rangle $ does not vanish. Actually,
if one chooses $\left\vert \chi_{R}(t_{f})\right\rangle =\left\vert
x_{R}\right\rangle $, we have $\Pi_{c}^{w}(t_{1})=K(x_{R},t_{f};w_{c}%
,t_{1})\psi_{2}(w_{c},t_{1})/\psi_{1}(x_{R},t_{f})$ that is seen to be
proportional to the propagator.\ As is obvious from Eq.  (\ref{amplitudelibre}%
), the fact that there are classical trajectories linking $w_{c}$ to $x_{R}$
with momentum $\tilde{p}=\left(  x_{R}-w_{c}\right)  /\left(  t_{f}%
-t_{1}\right)  $ is enough to guarantee the non-zero character of the weak
value, and hence the observation of a shifted pointer $c$. There are specific
choices of $\left\vert \chi_{R}(t_{f})\right\rangle $ for which $\Pi_{c}%
^{w}(t_{1})$ could vanish -- for example by imposing momentum filtering at
$x_{R}$; indeed if $\left\vert \chi_{R}(t_{f})\right\rangle $ is chosen to be
a Gaussian centered at $x_{R}$ with a mean momentum $p_{0}$ with a momentum
distribution sufficiently narrow so as to exclude  $\tilde{p}$, we will indeed
obtain $\int dx^{\prime}\chi_{R}^{\ast}(x^{\prime},t_{f})K(x^{\prime}%
,t_{f};w_{c},t_{1})\simeq0$. But except for these specific choices, we see
that we cannot say that the particle detected at $D_{R}$ came from the
wavepacket $\psi_{1}$ coming from the left -- given that the pointer $c$
shifts when turned on at time $t_{1}$. The difference with the single
wavepacket case studied of Sec. \ref{sec-2} is that here post-selection is
chosen at $D_{R}$ (thanks to $\psi_{1}$), whereas post-selection at $D_L$ was
suppressed (up to order $g^{2}$) in the single wave-packet case. The paths
contributing to this type of weak value are computed in Fig. \ref{fig-wt}(b).

Let us now focus on the pointer $b$ placed at the center. The interaction is
turned on at $t_{2}$, and the weak value reads%
\begin{equation}
\Pi_{b}^{w}(t_{2})=\sqrt{2}\frac{\left\langle \chi_{R}(t_{f})\right\vert
\hat{U}(t_{f},t_{2})\left\vert w_{b}\right\rangle \left\langle w_{b}%
\right\vert \left.  \psi(t_{2})\right\rangle }{\left\langle \chi_{R}%
(t_{f})\right\vert \left.  \psi_{1}(t_{f})\right\rangle }.\label{typ3}%
\end{equation}
Since the wavepackets interfere at $w_{b}$, the weak interaction couples
contributions of both wavepackets to the pointer, as evidenced by the
numerator.\ The corresponding paths are computed in Fig. \ref{fig-wt}(c). Typically $\Pi_{b}^{w}(t_{2})$ is non-zero, and the pointer shift
can therefore be observed.

Let us finally combine the different couplings we can make with our 3 weak
pointers.\ At $t_{1}$ we turn on the coupling for the pointers $a$ and
$c$.\ Both pointers will detect the coupling, due to the interactions with
$\psi_{1}$ and $\psi_{2}$ respectively [see Eqs. (\ref{typ1}) and
(\ref{typ2})]. If we turn all the pointers at $t_{2}$ (or additional pointers
that would sit at $a$ and $c$, in order to preserve the quantum state of the
pointers that had previously interacted at $t_{1}$), only the pointer $b$
interacts with the system, the pointer shift being proprotional to $\Pi
_{b}^{w}(t_{2})$ [Eq. (\ref{typ3})].\ At $t_{3}$ again both $a$ and $c$
interact with the system $\Pi_{a}^{w}(t_{3}),$ involving the interaction with
$\psi_{2}$ is now of the type given by Eq. (\ref{typ2}) and $\Pi_{c}^{w}%
(t_{3})$ similar to Eq. (\ref{typ1}). After post-selection at $D_{R}$,
combining the readings from all the weak pointers gives a view of the
superposition of paths emanating both from the left $(-x_{0})$ and the right
$(x_{0})$. These results call for a dynamical interpretation.

\section{Inferring the dynamics from the weak values\label{sec-inferring}}

\subsection{Weak trajectories: Local dynamics along Feynman paths}

A particularly simple way of understanding the weak values determined in Sec. \ref{sec-simplified}
is to follow the
system evolution with Feynman paths. The weak value of the
projector $\Pi_{w}=\left\vert w\right\rangle \left\langle w\right\vert $ at
time $t_{w}$ depends \cite{matzkinPR} on the sum over paths connecting a point $x_{i}$ within
the initial wavefunction $\psi(x_{i},t=0)$ to $w$ in time $t_{w}$ and then
going from $w$ to a point $x_{f}$ of the final distribution $\left\vert
\chi_{L}(x_{f},t_{f})\right\vert ^{2}$ in time $t_{f}-t_{w}$. In our problem,
the paths here are those carrying the wave-packet $\psi_{1}$ from the left
towards $D_{R}$ interacting along the way with the pointers, as well as the
paths carrying the wave-packet $\psi_{2}$ that get reflected by the
interactions towards $D_{R}$. In terms of the analogy with the double-slit
problem, we see that although post-selection is only compatible with the
wave-packet $\psi_{1}$ having gone through slit 1, the weak values reflect
nevertheless the presence of the wave having gone through the other slit, not
only at the midpoint $b$ (with the corresponding weak value $\Pi_{b}^{w}%
(t_{2})$), but also at the other positions in which a weakly coupled pointer
is placed: the corresponding weak trajectory starts at slit 2, and is turned
back towards $D_{R}$ by the interaction.

Note that if there are no paths connecting $w$ to $x_{f}$ in time $t_{f}%
-t_{w}$ the weak value vanishes and the system does not leave any trace on the
pointer. This can never happen for weak values of type 1 such as $\Pi_{a}%
^{w}(t_{1})$ given by Eq. (\ref{typ1}) since \ post-selection ensures that
such paths exist.\ However for weak values of type 2\ such as \ $\Pi_{c}%
^{w}(t_{1})$, for which the interaction involves the wavepacket $\psi_{2}$
that is orthogonal to the post-selected one, it is always possible to choose
the post-selected state so that the weak values of type 2 vanish. Since
post-selection involves filtering, this can be done by finding a
post-selection condition that filters away the paths (here imposing a momentum
filtering function at post-selection by choosing an appropriate function
$\chi_{R}(t_{f})$ would be the obvious choice). Recall that from a general
standpoint, a zero weak value in a given region can mean different things
\cite{matzkinPR}: (i) absence of the particle (when the interaction term in the
classical Lagrangian vanishes in that region); (ii) a vanishing wavefunction
by destructive interference of Feynman paths in that region; (iii) the
wavefunction propagated by the paths affected by the interaction is orthogonal
to the post-selection state (either the wavefunction does not reach the
post-selection region or summing over the individual amplitudes results in a
destructive interference). The latter case is at play when momentum filtering
is imposed.

\subsection{Non-locality and modular momentum}

A different take involving weak values in the same simplified double-slit
problem was proposed recently by Aharonov et al \cite{aharonov2S}. The idea is
to explain the interference pattern by way of a particle having a definite
location but sensing non-locally the potential due to the other slit.
Non-locality appears here through the evolution of the modular momentum
operator, which depends both on the value of the potential at the particle
position and on the value of the potential at a translated position (the
position is translated by the reciprocal of the modulation \cite{modular},
here the distance between the slits, i.e. between the mean positions of
$\psi_{1}$ and $\psi_{2}$ at $t=0$). According to Ref.\ \cite{aharonov2S} this
can explain the interference pattern produced (in our notation) when $\psi
_{1}$ and $\psi_{2}$ overlap in the neighborhood of $b$ while at the same time
one knows the path the particle took throughout given the post-selection condition.

Note however that asserting that the particle took a given path on the basis
of post-selection only works for specific choices of post-selected states,
those for which the weak values of type 2 mentioned above vanish. Otherwise
the non-local dynamics must also account for the appearance of a non-zero weak
value of type 2, that is for the fact that $\Pi_{c}^{w}(t_{1})\neq0$ although
the $\psi_{1}$ that will be post-selected is at $w_{a}$ at that time.

\begin{figure}[ptb]
	\centering \includegraphics[scale=1.]{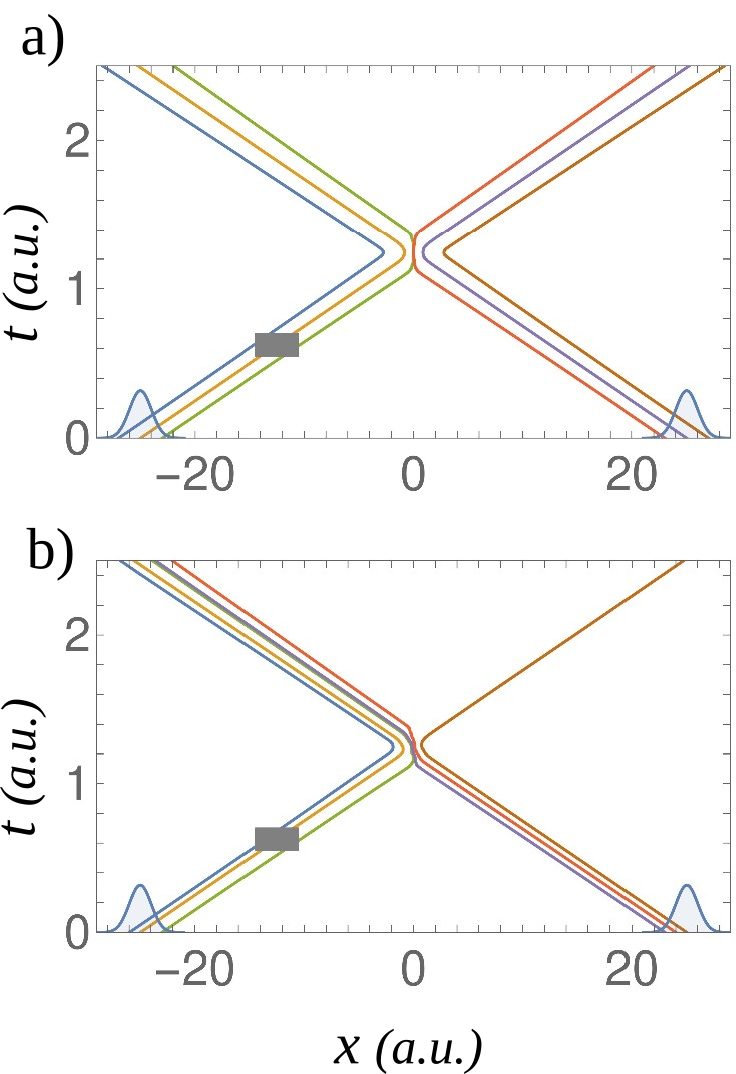}\caption{De Broglie-Bohm trajectories for the initial state given by Eq. (\ref{supp}) and an interaction with an interaction Hamiltonian given by Eq. (\ref{hint}) in a region centered at $w_a$. We use atomic units; the system has unit mass $m=1$, the pointer has a mass $M=10m$ and its initial state is  represented by a Gaussian centered at $y=w_a$.
		(a)  Weak coupling at $w_a$: the trajectories originating from $-x_0$ and $x_0$ and neighboring initial positions do not cross the midpoint, so that 
		post-selection at $D_R$ is due to the trajectories having started on the right, at $x_0$ (all these trajectories are computed with the pointer having initial position $y=w_a$ a.u.). (b) With a stronger interaction (though not yet in the strong regime), some trajectories are seen to cross (all the trajectories are computed with the pointer having initial position $y=w_a-2$ a.u.}%
	\label{fig-bohm-free}%
\end{figure}

\subsection{The de Broglie-Bohm picture}

The de Broglie Bohm (dBB) approach to the standard double slit problem is
well-known \cite{dewdney,holland}: a point-like particle is assumed to go
through one of the slits but its motion depends, through the quantum potential,while traveling along the current density arising
from the wavefunction going through both slits. In the one-dimensional analog
studied here, the dBB trajectories result from the interference between both
wavepackets. In the absence of any interaction, the trajectories do not cross, given the symmetric character of the wavepackets \cite{sanz}; the trajectories with initial position within $\psi_{1}$ evolve
first towards the right but as they approach the midpoint they are turned
back as the current density vanishes (due to $\psi_{1}$ and $\psi_{2}$
overlapping) and reverses its direction. This gives of course a dynamical interpretation
that is radically different than the one given in terms of weak trajectories,
since the post-selected particles at $D_{R}$ come from the right, not from the left.

If we now include interacting pointers, the numerical computation of
Bohmian trajectories becomes more involved. We have computed in
Fig.\ \ref{fig-bohm-free}(a)-(b) the resulting trajectories when only the pointer
$a$ is in place, and is turned on at $t_{1}$; we have taken
the pointer wavefunction $ \phi_{i}(y) $ to be given by 
a Gaussian (the computational details, based
on an expansion of free standing waves in a box \cite{matzkinSB} will be given
elsewhere). For a weak interaction Hamiltonian the dBB trajectories do not cross and are actually very similar to those in the
no-interaction case -- a consequence of the weakness of the coupling.  For significantly stronger couplings some trajectories do cross. An example is shown in Fig. \ref{fig-bohm-free}(b).

This paradoxical situation is reminiscent of other cases of mismatch between
the intuitively expected dynamics and the alleged motion of the Bohmian
particle (recall e.g., the case of surreal trajectories
\cite{surreal,vaidman-bullet}, or the persistence of non-classical
trajectories in semiclassical systems \cite{mismatch}). The model is
nevertheless consistent if one assumes that the weak interaction with the
quantum pointers does not involve the particle but only the pilot wave -- the
particle is only required to make detectors click, hence for strong
measurements. If the interaction is made sufficiently strong so as to acquire
which way information, then the Bohmian particle must interact with the pointer (given
that the particle can be detected), but  in the case of a
strong interaction the Bohmian trajectories initially associated with
$\psi_{1}$ do cross and are detected by $D_{R}$, a behavior that is quite
general \cite{hiley-surreal,laloe}.

\section{Conclusion \label{sec-conc}}

We have studied the dynamics that can be inferred from weak values in a simplified double-slit setup. While this
simplified setup does not capture the entire set of issues that appear in full double-slit problem, it does bring to the fore the which-path question. Positioning weakly coupled pointers between the slits and the detection screen induces additional constraints on the dynamical picture, as 
the pointer shifts -- given by the weak values -- can in principle be measured. Nevertheless, these constraints are not sufficient to pin down unambiguously the dynamical evolution of the system. The choice between admitting a local picture with a delocalized superposition, or a localized particle subjected to non-local interactions remains open.  It would be interesting to examine if, by considering more complex setups, some dynamical interpretations could be, if not ruled out impossible, rendered quite implausible.

\end{document}